# Spin-Chemistry concepts for Spintronics scientists
Review

Konstantin L. Ivanov,[1,2] Alexander Wagenpfahl,[3] Carsten Deibel,[3] Jörg Matysik[4,*]

[1]*International Tomography Center, Siberian Branch of Russian Academy of Science, Institutskaya 3a, Novosibirsk, 630090 Russia*
[2]*Novosibirsk State University, Institutskaya 3a, Novosibirsk, 630090 Russia*
[3]*Institut für Physik, Technische Universität Chemnitz, 09126 Chemnitz, Germany*
[4]*Universität Leipzig, Institut für Analytische Chemie, Linnéstr. 3, D-04103 Leipzig, Germany*

* Corresponding author, e-mail: joerg.matysik@uni-leipzig.de

**Abstract (max 350 Words)**

Spin-Chemistry and Spintronics developed independently and developed different terminology. Until now, the interaction between the two fields was very limited. Here we compile the two "languages" to enhance communication. We expect that knowledge of Spin-Chemistry will accelerate progress in Spintronics.

**Keywords (max 5)**
Radical pairs; triplet states; magnetic field effects; CIDEP; photo-CIDNP

**Introduction**
In general, chemical reactions are discussed in terms of thermodynamics: reaction enthalpy, reaction entropy and free energy. It is also recognized that steric and charge effects can lead to kinetic control of the reaction dynamics by introduction of activation energies. In some cases, chemical reactions are controlled by diffusional transport of highly reactive particles, e.g., free radicals, to the reaction zone. This view on chemistry is sufficient for processes which are spin-conserving, i.e., the spin multiplicity is not changed during the entire process. If during the course of reaction the spin-multiplicity is changed, spin-rules apply and Magnetic Field Effects (MFE), Magnetic Isotope Effects (MIE) as well as electron and nuclear spin polarizations might occur. This is the field of Spin-Chemistry [1-3].

The field of Spin-Chemistry emerged with the discovery of anomalous EPR (Electron Paramagnetic Resonance) intensities in $CH_4$ gas under irradiation by Fessenden and Schuler in 1963 [4]. Soon later, Bargon and Fischer observed anomalous NMR (Nuclear Magnetic Resonance) intensities upon thermal radical-pair formation [5]. Such anomalous intensity patterns are nowadays interpreted in terms of transient non-Boltzmann magnetization and called CIDEP (chemically induced dynamic electron polarization) and (photo-)CIDNP (chemically induced dynamic nuclear polarization). The theoretical description by Kaptein and Oosterhoff [6] as well as by Closs [7] in 1969 established Spin-Chemistry as a new field, initially mainly run by physical organic chemists as well as EPR and NMR spectroscopists. The discovery of CIDEP and CIDNP was followed by reports of MFE [8-9] on chemical reactions and MIE [10-11]. All these effects originate from the spin-conserving nature of most chemical reactions and from singlet-triplet interconversion in radical pairs, which is sensitive to external magnetic fields and local hyperfine fields of magnetic nuclei. Although experiments have been done in gas phase (see Sections IV.A and V.A of Ref. [2] and references therein) and in solid state (e.g., in photosynthetic reaction centers [12-14] and in organic solids [15-16]), Spin-Chemistry mostly deals with small organic molecules in the solution state. The radical-pair formation is now-a-days often caused photochemically either by electron-transfer or by bond-break, although also thermal bond-breaks cause radical pairs. With the introduction of the concept of the „spin-correlated radical pair" (SCRP) [17-20], Spin-Chemistry appeared to be a „completed field", in the notation of Heisenberg's *abgeschlossene Theorie*. Techniques of Spin-Chemistry, such as CIDNP, CIDEP, MARY (Magnetically Affected Reaction yield) and RYDMR (Reaction Yield Detected Magnetic Resonance) allow one to detect elusive paramagnetic species, such as radicals, radical pairs and triplet states, and to obtain their EPR parameters.

This review is dedicated to scientists of the field of Spintronics as an introduction into the older field of Spin-Chemistry. It seems to be economically reasonable to learn Spin-Chemistry language and concepts to prevent "re-invention" of previous knowledge. We will first review the „languages", which developed mostly independently in both fields. By providing a "dictionary", we can short-cut the introduction into the world of Spin-Chemistry for scientists from the Spintronics area. While, however, Spintronics focuses on spin transport, Spin-Chemistry is dealing with spin effects during chemical reactions. Since in this paper we use many acronyms, we additionally explain them all in a separate section.

| Table 1: Comparison of terminology, states | | |
|---|---|---|
| **State** | **Spin-Chemistry** | **Spintronics** |
|  | *Intermediate, transient* | *Topological excitation, defect* |
| R· | Spin states „$\alpha$" and „$\beta$" | Spin states „up" and „down" |
| R· | (Mobile) radical | Soliton |
| R$^-$/R·$^-$ | Anion/radical anion | Negatively charged polaron |
| R$^+$/R·$^+$ | Cation/radical cation | Positively charged polaron |
| R* (local) | Electronic excitation | Electronic excitation, exciton-bipolaron formation |
| R* (in crystal) | Electronic excitation | Exciton |
| R·$^+$-R·$^-$ | Radical-pair, SCRP | Radical-pair, bipolaron, polaron pair, charge-transfer exciton, (bound) electron-hole pair, geminate pair |
| $^3$R·· | Molecular triplet state | Triplet polaron pair, triplet exciton |
| R·-R· | Biradical | Soliton-antisoliton pair |

# 1. The language of Spin-Chemistry

The following two Tables compare the terminology of Spin-Chemistry and Spintronics for states (Table 1) and processes (Table 2). Obviously, two different "languages" have developed in parallel without having had much influence on each other, except probably by EPR spectroscopists. To ease communication between different scientific communities (and to avoid the situation that important concepts are lost in translation) we present Tables 1 and 2 as a simple "translator" between the two languages.

| Table 2: Comparison of terminology, processes ||| 
|---|---|---|
| **Process** | **Spin-Chemistry** | **Spintronics** |
| | *Spin-dynamics* | *Soliton and polaron dynamics* |
| 2R ➔ R·⁺-R·⁻ | Charge-separation, radical pair formation, electron transfer | Charge transfer, Charge-transfer exciton formation |
| ¹(R·⁺-R·⁻) ⇔ ³(R·⁺-R·⁻) | Singlet-triplet interconversion | Intersystem crossing[1], Singlet-triplet interconversion |
| ¹(R·⁺-R·⁻) ⇔ ³(R·⁺-R·⁻) | Phase coherence, quantum beats | Phase coherence, quantum beats |
| ¹(R·⁺-R·⁻) ⇔ ³(R·⁺-R·⁻) | Dephasing, $T_2$ relaxation | Dephasing, $T_2$ relaxation |
| ¹(R + R)* ➔ (³P + ³P) | Singlet fission | Singlet fission |
| ³R + ³R ➔ R + R | Triplet-triplet annihilation | Triplet-triplet annihilation |
| R· + R· ➔ R-R, <br> R·⁺-R·⁻ ➔ R-R | Recombination reaction | Soliton-antisoliton annihilation, charge-carrier recombination, geminate recombination |
| R·⁺-R·⁻ ➔ R·⁺ + R·⁻ | Escape reaction | Spin-diffusion |
| **R₁** + R₂ ➔ R₁ + **R₂** | Spin-diffusion | Polarization transfer |

Obviously the states are named very differently (Table 1). Even the two Zeeman states of a radical are often labeled differently. Molecules with an unpaired electron are called "radicals" (R·). Two radicals on the same molecule form a "biradical" (R·-R·). In this case, normally the two radicals are at the two ends of a long molecule. If the two radicals are on the same molecule and close together, they need to be in different orbitals, mostly one is in the highest occupied and the other in the lowest unoccupied molecular orbital. In this case, a molecular triplet state can occur (³R··). A pair of two radicals on two different molecules is called a "radical pair" (R·+ R·). This pair is often formed by the same chemical process, for example a bond-break or a photochemically induced electron transfer: in this situation it is formed in a particular spin state, either the singlet or the triplet state. Such a radical pair is termed a spin-correlated radical pair (SCRP). In liquids, where radicals can diffuse, radical pairs are usually classified as geminate pairs (G-pairs), i.e., pairs of radicals born in the same chemical event, or radical pairs formed upon encounters of free radicals in the solvent bulk (F-pairs).

---

[1] In Photochemistry, the term "inter-system crossing" (ISC) is only defined for an intra-molecular process (The recommendations of the IUPAC of terms used in Photochemistry, edited by S.E. Braslavsky, have been published in Pure and Applied Chemistry 2007, 79, 293-465). Therefore, in Spin-Chemistry, the change of spin-multiplicity, which occurs in an inter-molecular process, is termed "singlet-triplet interconversion". We would recommend using this term also in Spintronics.

For the more complex processes (Table 2), sometimes the same terminology is used. Interestingly, both spin-phenomena and processes are not only termed differently but also interpreted differently: A spin-chemist will discuss a radical-pair mainly under the aspect of its spin-evolution driven by internal interactions but will tend to ignore interactions with the environment. A spin-physicist, however, will often focus on electric polarization effects on the surrounding and might skip the magnetic forces between the two centers. One should note, however, that chemists are aware of the importance of electric polarization in chemical processes; a prominent example of theoretical understanding of electric polarization effects is given the famous Marcus theory [21] of electron transfer. Spin-orbit coupling is not a very prominent issue in Spin-Chemistry of radical pairs[2] because of the absence of heavy atoms in most (but not all) of the molecules, for which spin-chemical effects have been studied. For spin-physicists, however, it might be crucial since it can occur at defects determining how fast a triplet can go back to the ground state. On the other hand, spin-physicists often have been neglecting hyperfine coupling (HFC) interactions, which are a central issue in Spin-Chemistry. Furthermore, similar (or even identical) methods are termed differently.

| Table 3: Comparison of terminology, techniques | | |
|---|---|---|
| **Technique** | **Spin-Chemistry** | **Spintronics** |
| | *Reaction yield, reaction rate* | *Conductivity or resistance* |
| MFE detection | MARY, MFE | OMAR |
| Effect of resonance fields | RYDMR | EDMR, ODMR |
| Electron spin polarization | CIDEP, CIDEP in SCRPs | EPR of hyperpolarized charge transfer complex |
| Nuclear spin polarization | CIDNP | No analog |

Finally, Table 3 is dealing with existing methods. One can readily see that some methods are existing in both reviewed fields. It is obvious that both fields will profit from a fruitful exchange of ideas and concepts. In such a situation a better communication between scientists of the two fields is desirable not only for having a unified terminology, but mostly to avoid possible re-discoveries of the same methods. Using some examples we will show that scientists from both fields can learn from each other. The CIDNP method, despite its utility for studying short-lived radicals, has yet no analogue in Spintronics.

**2. Molecular systems**
Open-shell compounds, such as radicals, radical pairs and triplet states are in the heart of Spin-Chemistry. Radicals are frequent intermediates of many light-induced processes in chemistry. Furthermore, some stable radicals are known, which are not transient short-lived species but rather long-lived molecules with an unpaired electron. Radical pairs can be generated in various media by bond cleavage of a photo-excited molecule or by electron transfer from an excited electron donor to an acceptor (or by electron transfer to an excited electron acceptor from a donor). Such radical pairs inherit the spin state of their precursor. Likewise, biradicals can be formed by photo-induced intramolecular electron transfer or by bond cleavage in a cyclic molecule. Since the ground state of most molecules is the singlet state, triplet states are usually only transient species, which are formed upon light excitation with subsequent intersystem crossing producing the triplet state.

---

[2] Spin-orbit coupling is of importance for the triplet mechanism in Spin Chemistry and also for triplet state ONP and OEP; these cases are also briefly discussed in the text. One should also note that the difference, $\Delta g$, in g-factors of radicals, which is of great importance for Spin Chemistry, is also due to perturbation terms in the spin Hamiltonian coming from spin-orbit coupling.

For similar reasons magnetic and spin effects are of importance for physicists working in the Spintronics field. The corresponding devices allow the manipulation or detection of spins [22-23]. The most prominent type of organic Spintronics device is the spin valve, in which a thin organic semiconductor layer is sandwiched between two ferromagnetic electrodes [24]. A spin polarized current is injected from one of these electrodes and transported through the semiconductor. Another type also implements spin current, but without charge current, in a structure ferromagnetic metal/organic semiconductor/nonmagnetic metal, in which the first interface induce spin pumping [25]. Other devices do not rely on spin manipulation by ferromagnetic layers, but on the intrinsic properties of the organic semiconductor to show, for instance, organic magnetoresistance [26]. These devices also allow spin detection by electrical means. Organic light emitting diodes also rely on spin manipulation of radical pairs created from injected charge carriers in order to increase their electroluminescence quantum efficiency. Another molecular system in which spins play a major role are organic solar cells (OSCs) [27-28]. We choose it to serve as example, as it closely resembles the functional principles of systems found in Spin Chemistry. While OSCs have significant potential to become an inexpensive, large area and flexible photovoltaic technology at lower cost than conventional technologies, we will focus here on the spin processes from absorption to the generation of free charge carriers. The study of spins in organic semiconductors has a long-standing history, but their role in the fundamental processes in OSC has only very recently been highlighted in key publications [29-31]. Also, exploiting the unique properties of electronic spin interactions, the development of novel routes to enhance both the power conversion efficiency and lifespan of solar cells should be possible.

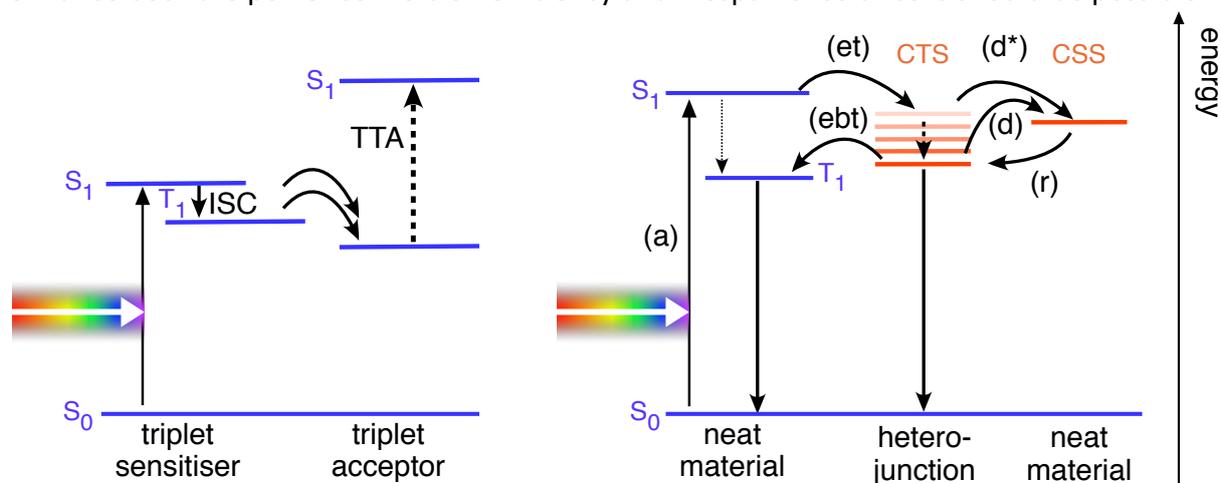

**Figure 1**. Spin dependent processes in organic solar cells. (Right) The steps from light absorption (a) towards generation of free charge carriers (d), potentially
supported by (Left) triplet–triplet annihilation (TTA), are described in the text. Only the lowest excited states of each kind are shown for clarity.

State-of-the-art OSCs consist of the combination of two organic semiconductors, (electron) donor and (electron) acceptor, as photoactive layer. These two materials are combined either as blend—the resulting architecture is called bulk heterojunction solar cell–or as two adjacent layers—yielding a planar heterojunction solar cell. The key processes for photovoltaic energy conversion in these types of OSCs are shown in Fig. 1. Ideally, singlet excitons in either donor or acceptor material are generated upon light absorption (a), here shown as $S_0 \rightarrow S_1$ transition. For sake of simplicity, we just consider the absorption taking place in the donor: the singlet exciton can diffuse towards the heterojunction, where an ultrafast electron transfer (et) to the acceptor occurs on the femtosecond time scale with almost unity yield. One reason is that this process is much faster than the intersystem crossing within, e.g., the donor from $S_1$ to the triplet state $T_1$. The resulting polaron pair, the negative polaron on the acceptor molecules with the positive polaron remaining on the donor molecules, is called "Charge Transfer State" (CTS) or "charge transfer complex". They have been reported to show the emission-absorption signatures in transient EPR as expected for SCRP [32]. The free charge carrier photogeneration in OSCs is mainly determined by the properties of the CTS. The dominant fraction of the CTS thermalizes [33] ENREF_28; only a small fraction might remain „hot" before dissociation [34-35] (d and d*, respectively) into free electrons and holes (CSS = Charge Separated State). These separated charge carriers can be extracted to yield the photocurrent. The role of the spin in several loss mechanisms [30, 36-37], which reduce the power conversion efficiency, is only partly understood. For instance, if the delocalisation of charge carriers in the CTS is limited, e.g., by energetic disorder, CTS dissociation will be uncompetitive compared to geminate recombination, leading to a lower photogeneration yield. The CTS can be in singlet and triplet configuration, although—due to the weak interaction within the polaron pair—energetically close. Therefore, the interconversion from singlet to triplet within the CTS might be comparatively fast. The detailed role of the spin in the geminate recombination (in competition with the charge photogeneration) is still unresolved. In principle, while the CTS singlet can recombine to the ground state, a spin flip (interconversion) to the CTS triplet can occur, which makes an electron back transfer ("ebt" in Fig. 1) into an intramolecular triplet state of either donor or acceptor possible. The loss in photocurrent due to the electron back transfer is not known quantitatively. It can be minimized by increasing the donor LUMO–acceptor LUMO gap [31], which shifts the CTS below the neat materials' triplet energies, but this tuning of the energy levels limits the achievable open circuit voltage. Up to now, it is unclear what the spin statistics of the interfacial CTS recombination are and how they are influenced by morphology and energetic or spatial disorder [38].

Two optimization strategies for improving the power conversion efficiency based on spin processes are singlet fission and triplet–triplet annihilation (TTA). Singlet fission is the spin allowed conversion of one spin singlet exciton to two spin triplet excitons, which can occur with high yield in some organic semiconductors.[39] It can therefore be seen as down conversion and so-called multi exciton generation. An enhanced power conversion efficiency is then foreseen, with the premise that fission of these high-energy singlet excitations into two independent triplets is quantitative, and that the resulting triplets subsequently dissociate into pairs of free charge carriers. The singlet fission process is also known to spin chemists; furthermore, it is known that this process is magnetic field-dependent [40-41]. Another approach for harvesting low energy photons is TTA [42] (see Figure 1). A singlet exciton is photogenerated in the triplet sensitizer molecule, and converted to a triplet by intersystem crossing (ISC). When two of such triplet excitons are transferred to the triplet acceptor, they can undergo TTA to generate a higher energy singlet exciton. The latter can be harvested by the solar cell concept described above. This concept allows, therefore, internal up-conversion of incident photons, thus extending the absorption range of the photovoltaic system to the little exploited near-infrared regime of the solar spectrum.

## 3. Spin-dynamics in Radical pairs

Radical-pairs, to be more precise, SCRPs allow for magnetic-field dependent chemistry. The Radical Pair Mechanism (RPM) is seen as the key mechanism[3] for magnetic field effects on chemical reactivity. The RPM attributes MFEs to (i) spin-selective recombination of radical pairs and (ii) to singlet-triplet interconversion, which is, in turn, sensitive to magnetic fields. Sometimes textbooks of organic chemistry state that the recombination reaction of the two radicals R$_1$• + R$_2$• ➔ R$_1$-R$_2$ typically occurs without any activation barrier. That statement, however, is only true if the radical pair (R$_1$• + R$_2$•) forms a singlet state, i.e., the electronic spin wavefunction is anti-symmetric. If the spin wavefunction is symmetric, i.e., the radical pair is in a triplet state, then the recombination is usually forbidden. Generally, the total spin of reactants should be the same as that of the reaction products:

$$\sum_{i,\text{reactant}} \hat{\mathbf{S}}_i = \sum_{k,\text{product}} \hat{\mathbf{S}}_k$$

meaning that the singlet radical pair can only recombine to a product in the singlet state (here summation is taken over each *i*-th reactant and *k*-th reaction product). Likewise, the triplet radical pair can only recombine to a product in the triplet spin state. In most cases, the two rates are considerably different with the singlet-state recombination usually being more efficient (although cases of more efficient triplet-state recombination are also known [43] and even should not be treated as exceptional). Spin-rules apply strictly and impose a rigorous kinetic control over thermodynamics. Hence, only singlet-state radical pairs recombine, while triplet-state radical pairs do not recombine, even when this would be energetically very favorable, and will follow an alternative reaction pathway.

Two radicals forming a radical pair can exist in four possible spin states. Namely, these states are a single singlet state $(\alpha\beta - \beta\alpha)/\sqrt{2}$, also called $S$ (or, sometimes, $S_0$) state, and three triplet states: $\alpha\alpha$, $\beta\beta$ as well as $(\alpha\beta + \beta\alpha)/\sqrt{2}$. The three triplet states are also called $T_+$, $T_-$ and $T_0$, respectively. Therefore, in three of four cases, a recombination of radicals, although thermodynamically favorable, is spin-forbidden. In this case, radicals would for example move apart or react with neighboring molecules to form more stable radicals. The latter reaction is spin-allowed because the total spin-state is not changed.

When a radical pair is formed in a particular spin state, singlet or triplet, its fate is different: in the former case fast recombination occurs, whereas in the latter case the radical pair decays through a different pathway. Because of this, the radical pair reactivity strongly depends on the rate of singlet-triplet interconversion. The strongest effects of such interconversion on the recombination yield are expected in the situation where the radical pair is born in a non-reactive state, thus, recombination can *only* occur after the interconversion takes place. In turn, the interconversion rate depends on the magnetic fields, which are external fields (static or oscillating) and the local fields of magnetic nuclei of radicals. Below we explain the origin of such a dependence and discuss its consequences. These consequences are the magnetic and spin phenomena in chemistry.

---

[3] One should note that RPM is not the only mechanism, since, e.g., the d-type triplet mechanism and triplet-radical mechanism, discussed below, also lead to magnetic field effects in chemistry.

In order to describe the spin dynamics of radical pairs on the quantitative level, the Stochastic Liouville Equation for its spin density matrix, $\hat{\rho}$, is commonly used [44-45]. This equation takes into account the coherent spin evolution (driven by the radical pairs Hamiltonian, $\hat{\mathcal{H}}$), spin relaxation and chemical reactions. Additionally, one can take into account the relative motion, which is described by a corresponding operator $\hat{\mathcal{L}}$, for instance, for diffusing radicals $\hat{\mathcal{L}} = D\Delta_r$ with the reflecting boundary condition at closest approach (here $D$ is the relative diffusion coefficient, $\Delta_r$ is the Laplace operator with $r$ being the distance between the radicals). The equation for the density matrix takes the following form:

$$\frac{d\hat{\rho}(\mathbf{r},t)}{dt} = \hat{\mathcal{L}}\hat{\rho}(\mathbf{r},t) - \frac{i}{\hbar}[\hat{\mathcal{H}},\hat{\rho}(\mathbf{r},t)] - \hat{\hat{R}}\hat{\rho}(\mathbf{r},t) - \hat{\hat{\mathcal{K}}}\hat{\rho}(\mathbf{r},t)$$

where $[\dots,\dots]$ is the commutator, $\hat{\hat{R}}$ is the relaxation super-operator and the $\hat{\hat{\mathcal{K}}}$ super-operator stands for spin selective recombination. Here, for simplicity, we do not discuss relaxation effects. When the radical pair selectively recombines from the singlet state, $\hat{\hat{\mathcal{K}}}$ acts on the density matrix in the following way:

$$\hat{\hat{\mathcal{K}}}\hat{\rho}(\mathbf{r},t) = \frac{w_S(\mathbf{r})}{2}\{\hat{\mathcal{P}}_S,\hat{\rho}(\mathbf{r},t)\}$$

where $w_S(\mathbf{r})$ is the position-dependent recombination rate, $\hat{\mathcal{P}}_S$ is the projection operator for the $S_0$ state and $\{\dots,\dots\}$ stands for the anti-commutator. For a static radical pair the position-dependent rate $w_S(\mathbf{r})$ can be replaced simply by a constant rate $k_S$. Such a form of the reaction operator corresponds to the decay of the singlet-state population at a rate $k_S$, whereas the phase elements, singlet-triplet coherences, decay at $\frac{k_S}{2}$.[46-47] Recently, possible corrections to such a form of the operator were discussed [48] differing in the decay of the coherences, which is faster than $\frac{k_S}{2}$ in some models. The time-dependent rate of the product is given by the following quantity:

$$R(t) = k_S\rho_{SS}(t) = k_S\text{Tr}\{\hat{\mathcal{P}}_S\hat{\rho}(t)\}$$

or, when the reactivity is position-dependent, by the integral over spatial coordinates:

$$R(t) = \int d\mathbf{r}\, w_S(\mathbf{r})\rho_{SS}(\mathbf{r},t) = \int d\mathbf{r}\, w_S(\mathbf{r})\text{Tr}\{\hat{\mathcal{P}}_S\hat{\rho}(\mathbf{r},t)\}$$

To calculate the steady-state reaction yield one should perform integration over time from 0 to $\infty$. The reaction yield $Y$ is obtained by integration of $R(t)$ from zero to infinity:

$$Y(B_0) = \int_0^\infty R(t)dt$$

In this formula we stress that the yield is the function of the $B_0$ field strength. The reaction operator presented here and the method of calculating $R(t)$ is valid for singlet-state recombination and weak spin-orbit coupling.

The Hamiltonian of the radical pair typically takes into account the Zeeman interactions of spins with the external field $\mathbf{B}_0$ (hereafter, the static field directed parallel to the z-axis), HFCs and electronic exchange interaction. For simplicity, we consider only the case of isotropic liquids. In this situation the Hamiltonian takes the form (here written in the angular frequency units):

$$\hat{\mathcal{H}}(\mathbf{r}) = \omega_{1e}\hat{S}_{1z} + \omega_{2e}\hat{S}_{2z} + \sum_j a_j^{(1)}\left(\hat{\mathbf{S}}_1 \cdot \hat{\mathbf{I}}_j^{(1)}\right) + \sum_k a_k^{(2)}\left(\hat{\mathbf{S}}_2 \cdot \hat{\mathbf{I}}_k^{(2)}\right)$$

$$- 2J_{ex}(\mathbf{r})\left\{(\hat{\mathbf{S}}_1 \cdot \hat{\mathbf{S}}_2) + \frac{1}{2}\right\}$$

Here $\omega_{ie} = g_i\mu_B B_0$ are the electronic Zeeman interactions with $g_1$ and $g_2$ being the electronic $g$-factors ($\mu_B$ is the Bohr magneton), $\hat{\mathbf{S}}_1$ and $\hat{\mathbf{S}}_2$ are the electron spin operators. We assume that each radical has a set of magnetic nuclei with spins $\hat{\mathbf{I}}_j^{(1)}$ and $\hat{\mathbf{I}}_k^{(2)}$ (the superscript denotes the radical, to which the nuclei belong) and HFC constants $a_j^{(1)}$ and $a_k^{(2)}$. Finally, $J_{ex}(\mathbf{r})$ is the position-dependent exchange coupling (also giving rise to the $r$-dependence of the Hamiltonian). In the presence of a transverse microwave (MW) field, which is commonly used to affect the spin evolution of radical pairs or to detect its CIDEP spectrum, one should add the corresponding terms to the Hamiltonian. Such terms are generally time-dependent but typically vanish in the MW-rotating frame of reference. The nuclear Zeeman interaction is omitted in the expression for the Hamiltonian because in liquids it is usually irrelevant. The reason is that at low fields this interaction is way too small to affect the spin dynamics whereas at high fields the nuclear Zeeman interaction simply changes the splitting between the eigen-states of the Hamiltonian $\hat{\mathcal{H}}$ corresponding to the nuclear states $\alpha$ and $\beta$ and does not affect spin mixing. In solids, however, such states are mixed by the anisotropic parts of the HFCs, which thus become relevant as well as the electron-electron dipolar coupling. In liquids anisotropic interactions are averaged out by molecular motion.

To simplify the description, it is common to present the spin-state of SCRPs using a vector model (Fig. 2).[1-2, 18-19] The arrows shown on the cones are not static but considered to precess with their Larmor frequency around the central axis. In this diagram, all four states are distinguished by different quantum numbers. In the singlet state, the total spin is zero. In fact, both arrows point into opposite directions and their magnetism is cancelled, i.e., a singlet state is not magnetic and does not interact with external magnetic fields. The situation is different for the three triplet states. Here, the magnetism does not disappear, and a triplet state is able to interact with external magnetic fields. While for the singlet state and the $T_0 = (\alpha\beta + \beta\alpha)/\sqrt{2}$ triplet, the energies are not affected by external magnetic fields, the $\alpha\alpha$ triplet gets destabilized while the $\beta\beta$ triplet becomes stabilized. Hence, the transition energies of a triplet are affected by external magnetic fields (Fig. 3). For radical pairs, the transition between $T_+ = \alpha\alpha$ and $T_- = \beta\beta$ is considered to be a double-quantum transition and is forbidden by optical means as well as in magnetic resonance. Such triplet states do not only occur in radical pairs but also by intersystem-crossing (ISC) at a single molecule mostly having one free electron in the HOMO and the second one in the LUMO.

The vector model also assumes that the electron spins precess about their effective magnetic fields, which are given by the superposition of the external fields and the local fields [49-50]. The scheme allows one to understand in a simple way how interconversion in radical pairs is occurring. $S$-$T_0$ conversion proceeds due to different precession frequencies of the two electrons: when one electron spin precesses about the z-axis faster than the other one, the radical pair oscillates between the $S$ state and the $T_0$ state. If one of the spins rotates about the x-axis (or y-axis), transitions between $S$ and $T_\pm$ triplet states occurs. We will use this description to give a simple explanation of magnetic and spin phenomena in radical pairs. Of course, such a simplistic treatment is not always applicable, in particular, for quantitative assessment of magnetic phenomena in chemistry. The reason, for instance, is that depicting the singlet state by two anti-parallel arrows is an over-simplification, which does not take into account the rules of quantum mechanics. Nonetheless, the vector model provides a reasonable qualitative view on the spin dynamics.

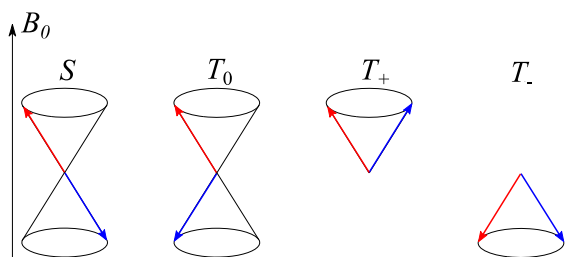

**Figure 2**. Vector model of the spin states of a radical pair. Here the red and blue arrows show the spin vectors of the two radicals; at high $B_0$ field each spin precesses on a cone about the $z$-axis, here $z||B_0$. The singlet state is the state with anti-parallel spins. In the $T_+$ and $T_-$ state there is positive and negative net polarization of the two spins on the direction of the $B_0$ field axis, respectively. In the $T_0$ state there is no *net* $z$-magnetization, but the total spin is non-zero.

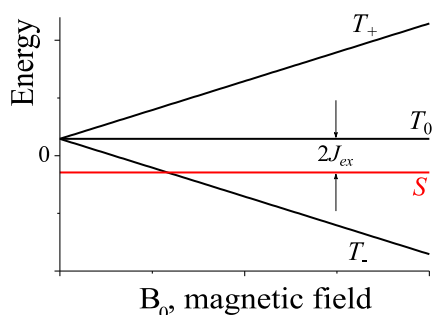

**Figure 3**. Schematic representation of the energy levels of a radical pair. The spacing between the $S$ and $T_0$ levels is equal to $2J_{ex}$. In radical pairs, typically, the singlet state is lower in energy at $B_0 = 0$; although the opposite situation can also be met. The splitting between the triplet levels linearly increases with the magnetic field due to the Zeeman effect.

### 4. MFE, MIE, MARY

As far as MFEs are concerned, their origin can be explained in simple terms using the vector model.[1-2, 18-19] To do so, we compare the situations of high external fields and low fields, as compared to the electron-nuclear HFC interaction. HFCs induce local magnetic fields in the $x, y, z$ directions in space, which can „rotate" the electron spins about the $x$-, $y$- and $z$-axes, see Fig. 4. At low fields, due to HFC all rotations are possible. Hence, all three interconversion pathways, $S \leftrightarrow T_-, S \leftrightarrow T_0, S \leftrightarrow T_+$, are operative. At high fields, however, the $S \leftrightarrow T_\pm$ transitions become energy forbidden because a small HFC cannot flip the electron spins. So, only the $S \leftrightarrow T_0$ conversion pathway is left, which is driven by the secular part of HFC and by the difference, $\Delta g = (g_1 - g_2)$ in the g-factors of the radicals. Consequently, the interconversion efficiency drops; in this simple model, roughly by a factor of 3. In real cases a quantum mechanical treatment should be used to calculate the conversion efficiency as a function of the external magnetic field strength. Thus, we obtain that the interconversion efficiency is sensitive to external magnetic fields giving rise to MFEs on chemical reactions. Such MFEs are well-established and can be found in a number of reactive systems. For further detail we recommend reviews on this subject.[1-3, 51-54]

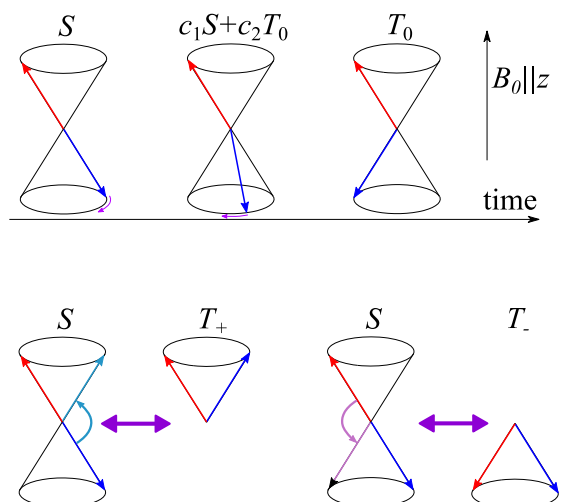

**Figure 4**. Schematic representation of singlet-triplet transitions in a radical pair. (top) $S$-$T_0$ transitions occur due to the difference in the precession frequency of the two radicals. Consequently, a radical pair starting from the singlet state transforms to a superposition of the $S$ and $T_0$ states, then to $T_0$ and back. The difference in the precession frequencies can be caused by secular HFCs and by $\Delta g \neq 0$. (bottom) Flips of the spins, e.g., due to local HFC fields or due to external MW-fields, can lead to mixing between the S state and $T_\pm$ states. The $B_0$ field is parallel to the $z$-axis.

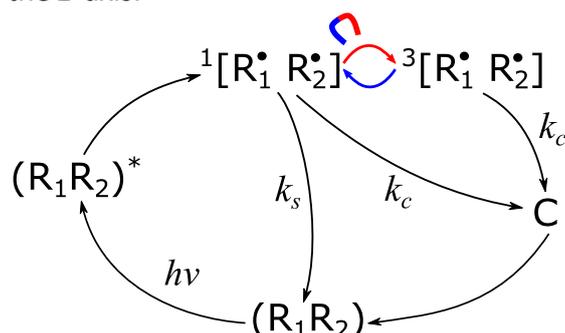

**Figure 5**. Formation of MFEs upon recombination of SCRPs. In this example the (R$_1$R$_2$) molecule goes to the singlet-excited state (R$_1$R$_2$)$^*$; subsequently, the SCRP [R$_1^\bullet$ R$_2^\bullet$] is formed in the singlet state. The singlet-state SCRP can recombine to the ground state or go to the triplet state by interconversion: this stage is magneto-sensitive as symbolically indicated. Both singlet and triplet SCRP can go to the state C, e.g., SCRP can undergo diffusional separation to form escaped radicals. Typically, the interconversion is slower at higher fields, i.e., formation of (R$_1$R$_2$) is more efficient at high fields. For the triplet precursor the MFE is the opposite: formation of (R$_1$R$_2$) is less efficient at high fields. The same scheme can be used to explain MIE: the interconversion rate depends on the content of magnetic isotopes in radicals. The interconversion rate is, e.g., higher for $^{13}$C enriched radicals. One should note the similarities with Fig. 1: (R$_1$R$_2$)$^*$ corresponds to (singlet) exciton in neat material: $^1$[R$_1^\bullet$ R$_2^\bullet$] and $^3$[R$_1^\bullet$ R$_2^\bullet$] correspond to CTS; (R$_1$R$_2$) = ground state; C = charge separated state, CSS.

The formation of MFEs on recombination of SCRPs can be explained using the scheme shown in Fig. 5 (the outline of this figure is following that of Fig.1 of Ref. [53]). Due to the presence of the magneto-sensitive interconversion stage the reaction yield becomes sensitive to the magnetic field strength. In this example, when the radical pair is born from a singlet precursor the yield of (R$_1$R$_2$) is higher at high magnetic fields. When the initial state of the radical pair is a triplet the MFE is just the opposite. The size of the MFE also depends on the precursor. Specifically, MFEs are stronger when the magneto-sensitive interconversion is the kinetic bottleneck of the process. In liquids, MFEs can be formed for recombination of G-pairs as well as for F-pairs. The formation of MFEs of G-pairs generated in a particular spin state is the same as described above. For F-pairs formed in a random spin state the existence of MFEs is, at first glance, puzzling. However, one should bear in mind that the size of MFEs for singlet-born and triplet-born is different, i.e., MFEs from such pairs, though opposite in sign, do not compensate each other completely.

For observing MFEs on chemical reactions one can either monitor the concentration of the reaction product in real time at various magnetic fields or perform a steady-state experiment. For observing MFEs in steady-state experiments, it is necessary to have some "branching", which makes the overall product yield dependent on the interconversion rate, i.e., on the external magnetic field strength. In the absence of such branching all radicals pairs would eventually recombine, leading to cancellation of potential MFEs. In the example given in Fig. 5 branching is provided by the reactions, in which C is formed.

MFEs can be obtained not only for recombining radicals but also in other cases where spin interconversion affects the reactivity. Such cases are the quenching of excited triplet states by radicals [55-56] and triplet-triplet annihilation [57-59]. In the former case, the process is usually allowed from the doublet state (producing the molecule in the singlet ground state and not causing any chemical changes of the radical) so that the magneto-sensitive doublet-quartet interconversion comes into play. In the latter case, the total spin of the reactants can be equal to zero (singlet), one (triplet) or two (quintet) with only the singlet reaction channel being reactive. This spin selectivity in combination with the magnetic field-dependent interconversion can give rise to MFEs.

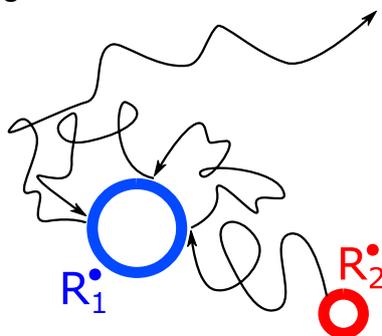

**Figure 6**. Re-encounters of radicals. In liquids particles usually move by means of diffusion. In this situation two radicals, $R_1$ and $R_2$, collide a few times before they escape to the solvent bulk and lose correlation with each other (after the characteristic time, which is equal to $\tau_D = R^2/D$).

The same scheme can be used to obtain a simple explanation of MIE. MIE in chemical reactions are not due to the difference in mass of isotopes (which is less than 10% for $^{12}$C and $^{13}$C) but due to the different spin and gyromagnetic ratio of them. For instance, the $^{12}$C carbon isotope is non-magnetic having zero spin, whereas $^{13}$C is a magnetic spin-1/2 nucleus. Consequently, in a $^{13}$C labelled radical HFC appears, which can drive the interconversion and make it faster. MIEs can be positive or negative depending on the initial state of the radical pair. Typical applications of MIEs are fractioning of isotopes and elucidating the mechanism of chemical reaction: the presence of MIE provides a clear evidence that radical pairs are reaction intermediates, and allows one to identify the spin multiplicity of reaction intermediates. For learning more about MIE in chemistry, we advise the reader to go for more specialized reviews [60-62].

In this context it is also important to mention the so-called „cage effect", which is crucial for MFEs in liquids.[1-3] Generally, the spin dynamics of radical pairs needs certain time to develop. The typical times required for interconversion are on the nanosecond timescale meaning that the radical pair partners need to stay close to each other for a time period of comparable duration. This becomes possible due to the cage affect: solvent molecules trap radicals and do not let them separate immediately. Importantly, when radicals undergo stochastic motion (diffusion) due to "kicks" from solvent molecules, they collide many times before radical pair separation, i.e., numerous re-encounters of radicals occur. Generally, for radicals separated by a distance $r$ the probability of at least one re-encounter is equal [63] to $R/r$. Consequently, the two radicals spend an extended time in the proximity of each other and completely loose correlation with each other after the characteristic time, which is equal [64] to $\tau_D = R^2/D$, see Figure 6. For spherical particles diffusing in three dimensions $R$ is the closest approach distance equal to the sum of the radical radii.

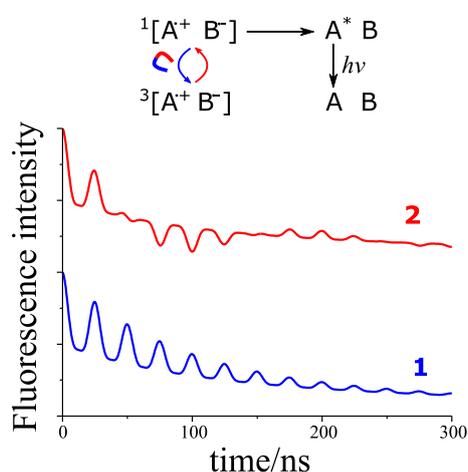

**Figure 7**. Calculated time-resolved MFE traces as obtained by monitoring recombination fluorescence, resulting from recombination of radical ion pairs in singlet state (such a situation is frequently met upon pulsed radiolysis of non-polar solutions of electron and hole acceptors). Magneto-sensitive interconversion (indicated in the reaction scheme) results in oscillations (quantum beats) of the singlet-state population, which manifest themselves in the fluorescence of A[*]. Quantum beats can be due to HFCs (curve 1, HFC-driven quantum beats) and also due to $\Delta g \neq 0$ (curve 2, HFC-driven and $\Delta g$-driven quantum beats are superimposed). For convenience of the reader trace 2 is shifted along the vertical axis (at $t = 0$ the fluorescence intensity is the same in both cases). Here we consider a radical ion pair with for equivalent magnetic nuclei with HFC of 20 MHz; for curve 2 the $\Delta g$-term is taken equal to 5 MHz. Fluorescence decays due to recombination of the SCRP (additionally, quantum beats decay due to relaxation).

MFEs can be studied using different techniques. Concentration of radicals, as well as the radical pair recombination yield can be traced by detecting optical absorption, luminescence, photo-current, etc. These quantities can be monitored either in steady-state experiments or in a time-resolved fashion. Time-resolved measurements can reveal very unusual behavior of the reaction rate of radical pair recombination: this rate has an oscillatory component due to the coherent nature of singlet-triplet mixing.[65-67] The oscillations, often termed „quantum beats", are driven by HFC and $\Delta g$. Hence, the frequency of the quantum beats is given by HFCs and, when the magnetic field is sufficiently strong, also by the $\Delta g$-term, see Figure 7. Time-resolved MFEs allow one to obtain EPR parameters of radical pairs, which are often too short-lived for detection by conventional EPR methods. Quantum beats are usually observed on the nanosecond timescale; however, when the $\Delta g$-term is extraordinarily large quantum beats occur on the picosecond timescale [68].

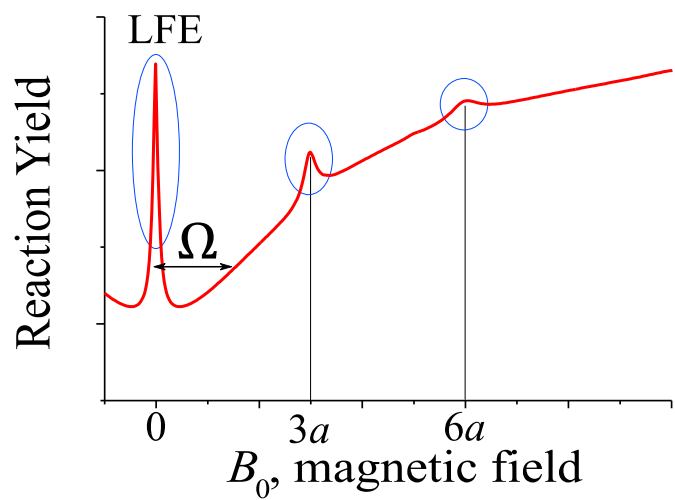

**Figure 8**. MARY curve, i.e., magnetic field dependence of the reaction yield for a singlet-born radical pair. Here sharp maxima correspond to crossings of the spin energy levels of the radical pair. A crossing at zero field gives rise to the Low-Field Effect (LFE). Additional level crossings are found at specific field strengths (e.g., for radical with 6 equivalent magnetic nuclei these fields are $3a$ and $6a$). The width, $\Omega$ (given by the effective HFC of the radical pair), of the broad MARY line at $B_0 = 0$ is also indicated.

The magnetic field dependence of the reaction yield of a radical pair is often termed a MARY curve. MARY curves typically contain maxima and minima with their positions depending on the EPR parameters of radical pairs, see Figure 8.[69-73] At zero field, there are two kinds of features observed: a broad feature having a width $\Omega = \sqrt{\frac{2}{3}\sum_k a_k^2 I_k(I_k + 1)}$ (summation is performed over all magnetic nuclei of the radical pair having HFC constants $a_k$ and spins $I_k$) of about the effective HFC in the radical pair and a sharp feature, often termed Low-Field Effect (LFE). The broad feature is, in fact, a level anti-crossing effect, whereas the sharp feature results from a pure level crossing, which is always present in radical pairs at zero field (originating from the equivalence of all directions in space). As usual, we assume that level crossing corresponds to a situation where two levels, $|K\rangle$ and $|L\rangle$, become degenerate at particular field strength. However, it is known that if there is small perturbation, $V_{KL}$ mixing the two levels they never cross. Hence, the level crossing turns into a level anti-crossing (often termed avoided crossing); at the level anti-crossing the initial states $|K\rangle$ and $|L\rangle$ become mixed. The width of the LFE feature is given by the inverse decoherence time in the radical pair, which is coming from the electronic spin relaxation as well as from chemical reactions, i.e., radical pair recombination and radical pair transformation. Hence, the LFE width can be used to determine rates of fast chemical processes on the nanosecond time scale. Interestingly, in some radical pair systems additional sharp features at $B_0 \neq 0$ can be found resulting from additional level crossings. For instance, in radical pairs comprising radicals of hexafluorobenzene (which has six equivalent $^{19}$F nuclei) there are sharp features on top of a smooth background have been found for $B = 0, B = 3a, B = 6a$; here $a$ is the $^{19}$F HFC constant of the hexafluorobenzene radical anion.[74-75] Such findings are in agreement with analytical theory, which can be developed for radicals with a set of equivalent nuclei [76]. Sharp features coming from level crossings (at $B_0 = 0$ as well as at non-zero fields) have been found [70-71, 74-75, 77] for a number of experimental systems where the SCRP has a set of equivalent nuclei.

Considerable MFEs and additional features in MARY curves can also arise due to other interactions, notably, due to the electron-electron exchange interaction. In the presence of $J_{ex}$ there is an energy gap between the singlet and triplet levels; consequently, the interconversion slows down. Only at a particular field strength, $B = 2|J_{ex}|$, which matches the singlet-triplet energy gap, one of the triplet levels, $T_+$ or $T_-$, tends to cross with the singlet level. At this field the interconversion becomes efficient due to the transitions between the crossing energy levels; these transitions are operative due to HFCs, which also turn the level crossing into an avoided crossing. Consequently, in the field dependence of the reaction yield a peak or a dip is observed at $B_0 = 2|J_{ex}|$. Additional features can appear at other matching conditions, for instance, when the HFC term matches the $\Delta g$-term: upon such a matching the interconversion in a particular nuclear spin ensemble is slowed down.

The MARY and OMAR (standing for organic magneto-resistance) [78] techniques represent, in fact, essentially the same method, despite having different names. The OMAR effect is observed by monitoring the resistance of an organic material originating from the spin-dependent nature of charge carrier transport and recombination. Essentially, the mechanism underlying OMAR is the same as for MARY. On the one hand, MARY is a more general term, since MARY is not supposed to be bound to only organic systems (examples of MARY in inorganic systems also exist [77]) and does not imply that resistance is used to monitor reaction yield. In fact, in OMAR, effects known for MARY have been reported. For instance, the LFE-type behavior has been found [79-81] in a number of systems used for Spintronics applications. On the other hand, OMAR is just one instance of MFE, which has been reported not only for (magneto)resistance, but also electroluminescence and other observables [82].

The exact mechanism leading to MFE, for instance observed in organic diodes by means of OMAR, is still not completely understood. Several relevant processes have been proposed: as mentioned above, they are all similar in relying on spin-selective reactions of particle pairs. The most important underlying mechanism is spin mixing of the particles by hyperfine interaction, which is suppressed by the magnetic field. The particle pairs are bipolarons, electron-hole (polaron) pairs (or charge transfer excitons), but also polaron interacting with triplets as well as triplet exciton pairs. The bipolaron model [83] can explain a positive magnetoresistance in energetically disordered systems, such as a conjugated polymer film. A mobile polaron is able to hop to a site already occupied by a polaron of the same charge type (unipolar OMAR), if the thus generated bipolaron is in the singlet state, while the triplet formation is unfavourable. The singlet bipolaron enhances current flow, wheres the triplet configuration blocks the current. This so-called spin blocking can be cancelled by the magnetic field. The spin mixing seems to be most pronounced in the slow hopping regime [84], for instance when deep traps reduce the charge carrier mobility [85]. Oppositely charged polaron pairs [16] (or correlated radical ion pairs) can show a bipolar OMAR, if the charge transport is limited by spin-selective electron-hole recombination [86]. In contrast to the unipolar OMAR where bipolarons enhance the current flow, here two (oppositely charged) polaron pairs usually reduce the current: either by recombination or generation of a triplet exciton. Accordingly, the resulting OMAR is usually negative [84], although the original model [82] can in principle accommodate also positive changes. The $\Delta g$-mechanism described above can show MFE [82], usually of the opposite sign than the OMAR due to suppression of the hyperfine induced spin mixing. Spin-orbit coupling is usually of lesser importance, unless the hyperfine interaction is strongly suppressed—e.g. in C60, which lacks the protons and contains 99% $^{12}$C, having zero nuclear spin [87]—or heavy metal atoms are present as part of the molecule or by doping [86, 88]. An MFE can also be observed by interactions with excitons: polaron–triplet interactions [89], triplet–triplet annihilation [90], or singlet fission [40].

Generally, the magnetic field is characterized not only by its strength but also by its direction with respect to the molecular axes system. Hence, there is not only the dependence of the reaction yield on the field strength, but also on the molecular orientation [91], for instance due to the anisotropy of the hyperfine interaction. The orientation dependence is of importance in solids, since in liquids molecules usually tumble so fast that anisotropic spin interactions are averaged out and all field directions in space become completely equivalent. However, in solids the reaction yield and the MFE can depend on the direction of the external magnetic field [91-93].

## 5. RYDMR

One more important member of the family of spin chemistry techniques is RYDMR.[94-97] The idea of RYDMR is based on affecting the singlet-triplet interconversion in radical pairs by applying resonant MW-fields, see Fig. 9. When such MW-fields enhance or slow down the interconversion, the reaction yield is altered. Of course, MW-fields affect the interconversion only when they are applied resonantly to some of the EPR transitions in the radical pair. Therefore, variation of the MW-frequency (to be more precise, by variation of the external magnetic field, as usually done in EPR) allows one to obtain the EPR spectrum of the radical pair by monitoring the reaction yield. This is the essence of the RYDMR technique, which can be used for two purposes: (i) controlling the reactivity of radical pairs by using spin degrees of freedom and (ii) obtaining EPR parameters of short-lived radicals and radical pairs.

RYDMR spectra can be obtained [95] by monitoring optical absorption, luminescence of the reaction product or photocurrent from radical ions, which escape recombination, i.e., RYDMR exists in different versions depending on the observable. Although RYDMR is not as generally applicable as EPR, it has advantages, namely, sensitivity and time resolution. In addition to the possibility of acquiring EPR spectra of radical pairs one can also perform more complicated experiments. For instance, it is possible to obtain quantum beats in the recombination efficiency, which are indicative of the coherent spin dynamics and can be used for precise measurements of EPR parameters.[98-99] Typically, selective excitation of one of the radicals results in quantum beats with the nutation frequency $\omega_1$ of the MW-field, whereas non-selective excitation (spin-locking) produces "double beats" with the frequency of $2\omega_1$. Somewhat later than in Spin Chemistry such effects were discussed [100-101] in the Spintronics field. One more important aspect of using RYDMR is that it allows one to affect the reactivity of radical pairs. In the easiest way this can be done by applying a strong MW-field, which drives the EPR transitions of both partners of the radical pair. In this situation „spin-locking" takes place: the singlet state of the two spins is isolated from the triplet state. Consequently, the interconversion is blocked (if a very strong field is used) or at least suppressed.

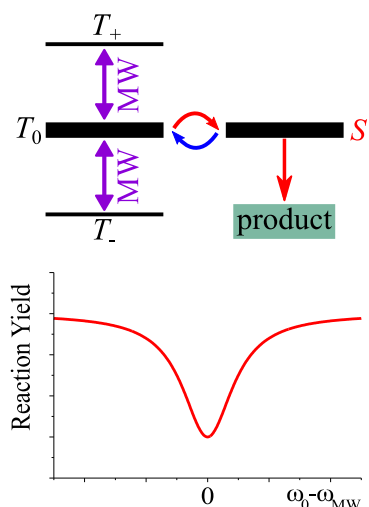

**Figure 9**. Principle of the RYDMR method. (top) Reaction scheme: interconversion mixes the $S$ and $T_0$ of a radical pair, which reacts from the $S$ state. Hereafter, in the energy level diagram the thickness of levels corresponds to the state population. A MW-field can drive the $T_0 \leftrightarrow T_+$ and $T_0 \leftrightarrow T_-$ transitions. (bottom) Schematic representation of the RYDMR spectrum: the singlet-state population can be monitored by measuring the reaction yield as a function of the MW-frequency. When the MW-field is applied in resonance with the triplet transitions, $T_0 \leftrightarrow T_+$ and $T_0 \leftrightarrow T_-$ (i.e., the MW frequency is matched to $\omega_0$, which is the precession frequency of the SCRP partners), the populations of the $S$ and $T_0$ states are decreased. The RYDMR signal can be obtained as a dip in the dependence of the reaction yield on the MW-frequency, $\omega_{MW}$. Experiments with MW-pulses with variable lengths and inter-pulse delays can be performed to obtain quantum beats and echo-like signals.

The RYDMR equivalent in Spintronics is EDMR or ODMR [16, 102]. Recently, important results have been obtained in the EDMR field [102]: pulsed EPR experiments have become feasible by combining EPR pulse sequences with sensitive current detection. In such experiments echo-type signals have been obtained. One should note that RYDMR and EPR use different observables; for this reason, one cannot use EPR pulse sequences in RYDMR without a pertinent modification [103]. Specifically, in EPR spin magnetization is detected, whereas the RYDMR signal is maximal when the radical pair is in the singlet state. However, in such a pair all magnetization components are zero. Likewise, pure spin magnetization does not give any contribution to the singlet state population because the singlet spin order is essentially a two-spin order.

These problems can be overcome by using a modification of spin-echo experiments in EPR [102]. At the instant of time where the spin echo is formed the refocused magnetization can be converted into two-spin order by applying an additional 90-degree pulse. The echo is then monitored by changing the instant of time when the last pulse is applied. Such a scheme can be modified further to exploit more advanced EPR methods for RYDMR purposes. For instance, the feasibility of the electron-electron double resonance [104] and electron-nuclear double resonance [105] has been demonstrated.

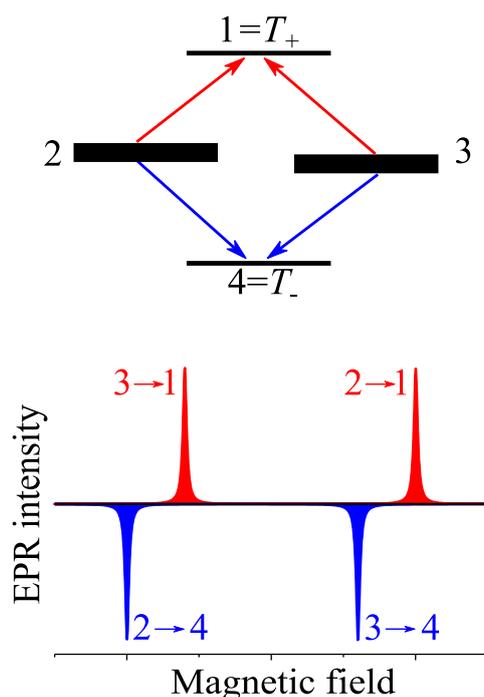

**Figure 10**. (top) Populations of the electronic spin state of an SCRP: when the radical pair is singlet-born initially only the two central states, $|2\rangle$ and $|3\rangle$ have "singlet character" (each of them is a superposition of $|S\rangle$ and $|T_0\rangle$), are populated, resulting in two EPR transitions in absorption and two in emission. Positions of the energy levels are out of scale. (bottom) Schematic representation of the corresponding CIDEP spectrum, consisting of two anti-phase doublets.

## 6. CIDEP and CIDNP

In this section, we provide a short description of spin "hyperpolarization" generated in chemical processes. Hyperpolarization refers to non-Boltzmann spin polarization which is highly desired by spectroscopists since it enhances the sensitivity of the method. Polarization of electron spins (CIDEP) and nuclear spins (CIDNP) results from spin selectivity of chemical reactions and can be used for sensitive detection of transient radical species. We start our description from CIDEP and introduce the main mechanisms for CIDNP formation.

In SCRPs, singlet and triplet states are not eigenstates but allowed to evolve in a coherent superposition of states. At high fields, in particular, a dynamic interconversion between the $S$ and $T_0$ states occurs if there is a difference in the precession frequency of the SCRP partners. There are two reasons for having different precession frequencies: (i) both individual radicals have different $g$-values, i.e., have different "chemical shifts" on their EPR axis which is due to different chemical environments. In this case $\Delta g$ is unequal to 0. (ii) Coupling of an electron spin to magnetic nuclei by HFC interaction will either accelerate or slow down the precession frequency of the electron depending on the direction of the nuclear spin state. As a consequence of the spin mixing the population becomes evenly distributed between the two spin states, $S$ and $T_0$.

SCRPs, in contrast to radical pairs with equilibrium populations of the spin states, show a particular intensity pattern in the EPR experiment (Figure 10).[1, 3, 106-107] While a thermally relaxed radical pair shows transitions with absorptive (positive) intensity, a SCRP born for example by bond-break from a singlet state would populate initially only the $S$ state, which has 50% αβ and 50% βα characteristics. This population distribution leads to a special pattern showing both emissive (negative) and enhanced absorptive (positive) signal intensities. The spectrum thus consists of two "anti-phase" doublets. Such transient electron spin order is often observed in EPR spectroscopy, corresponding to CIDEP. For observing such CIDEP spectra it is required that either $J_{ex}$ is non-zero or that a non-averaged electronic dipolar coupling is present, which is often the case in solids.

A more complex situation arises in solids where a distribution of different interactions exists, notably, of g-anisotropies and of the electronic dipolar coupling. As a result, the shape of the spectral doublets can change but the "anti-phase" nature of polarization remains.[106-107] The antiphase polarization originates from the spin-selective formation of radical pairs: in this situation no net polarization is expected (because the initial magnetization of the pair of spins is zero). Instead, two-spin order is formed. This spin order results in different phase of EPR lines within each multiplet; for this reason, such CIDEP is sometimes termed "multiplet CIDEP". Detection of spin echo from SCRPs is also possible by EPR methods. It is worth noting that the spin echo is collected 90 degrees "out-of-phase" [107-109]. This feature of the spin echo formation is markedly different from the spin echo coming from thermally polarized spins, allowing one selective detection of SCRPs. Furthermore, the out-of-phase echo signal is modulated due to the electronic spin-spin coupling, exchange or dipolar. Consequently, a precise determination of this coupling becomes feasible, providing the information about the SCRP structure, notably, about the distance between the radical centers.

In liquids, detection of CIDEP of SCRPs is often impossible because radicals quickly separate. Nevertheless, CIDEP of radicals can be often detected. Such CIDEP usually results in the opposite phase of polarization of the partner radicals avoiding geminate recombination: when the SCRP is singlet-born, no *net* spin polarization can be formed. The formation of such CIDEP can be explained using the fictitious spin representation of CIDEP, as proposed by Adrian [110].

Other CIDEP mechanisms are also known. CIDEP can be generated from molecular triplet states. The corresponding mechanism is termed "triplet mechanism" [1-3], see Section "Optical Nuclear Polarization". In this situation, the polarization formation is due to the difference in the ISC rates for different triplet substates in non-symmetric molecules. For instance, when a triplet $T_1$ state is formed from an excited singlet state, $S_1$, the ISC rate is different for the three triplet sublevels, $T_X, T_Y, T_Z$. Consequently, the triplet state is formed in a non-equilibrium spin state and exhibits CIDEP. This mechanism is termed the "p-type" (population-type) triplet mechanism [111-112]. Alternatively, CIDEP can be formed due to the "d-type" (depopulation-type) triplet mechanism [113] when the decay of the triplet state is different for the different sublevels. In contrast to the "p-type" triplet mechanism, "d-type" triplet mechanism can also lead to magnetic field effects on product yield [114-116]. When a radical pair is formed from the polarized triplet state, it inherits the triplet-state CIDEP. Finally, we would like to mention that CIDEP can be due spin-selective processes involving particles with higher spin, e.g., due to the radical-triplet pair mechanism [117-118]. CIDEP effects have already been observed in materials, which are used for OPV: Behrends et al. [32] and Kobori et al. [119] have detected anti-phase EPR lines of photo-induced charge transfer complexes, i.e., of SCRPs, whereas Lukina et al. [120] have recently reported a study of the out-of-phase electron spin echo.

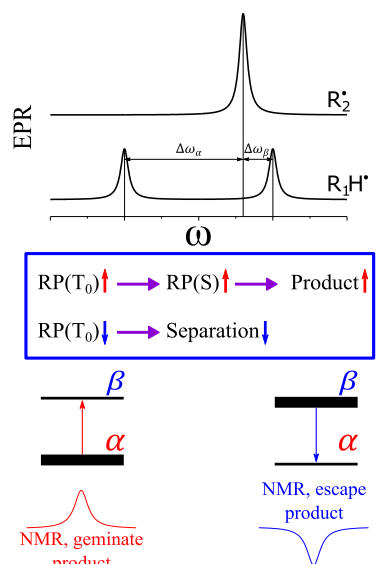

**Figure 11**. Scheme of CIDNP formation by spin sorting at high magnetic fields. (top) EPR spectra of the two radicals: one radical has a single spin-1/2 nucleus resulting in splitting of the EPR line into two components; the other radical has no nuclei, i.e., a single EPR line. The difference in EPR frequencies, $\Delta\omega_\alpha$ and $\Delta\omega_\beta$, are indicated for the radical pairs with the nucleus in the $\alpha$-state and $\beta$-state, respectively. Since $\Delta\omega_\alpha \neq \Delta\omega_\beta$, the $S$-$T_0$ interconversion rate is different for the two nuclear spin states. (middle) Kinetic scheme of CIDNP formation: a triplet-born radical pair in the $T_0$-state rapidly goes to the reactive single state for the $\alpha$ nuclear spin state, resulting in reaction product enriched in this nuclear spin state. For the $\beta$ nuclear spin state radical pairs react less efficiently and undergo separation. (bottom) The geminate reaction product is enriched in the $\alpha$ nuclear spin state giving an NMR line with enhanced absorption, whereas for reaction products of the escaped radicals opposite sign of polarization and opposite CIDNP sign is expected.

Until now, no CIDNP phenomenon [121-124] has been observed in spintronics, although a possibility of obtaining such effects has been mentioned [125]: "If nuclear spin resonance is found to have an impact on the spin-dependent electron transport due to the hyperfine interaction, ultimately the opposite process may become possible: storing electronic spin information in the nuclear spin." Despite that we want to describe the basic CIDNP theory, based on the classical RPM [1, 3] intending to stimulate future NMR research in that field. Fig. 11 explains the situation for a radical pair with a single proton (R$_1$H + R$_2$). Here we schematically show the EPR spectrum of the radical pair assuming that the two partners R$_1$H and R$_2$ have slightly different *g*-values. Therefore the two electrons have also slightly different precession frequencies and oscillate between $S$ and $T_0$ states. The R$_1$ signal in the EPR spectrum in Fig. 11, however, is split into two components. The origin of this split is due to the HFC of the nearby proton. Also the proton has two nuclear spin states, either $\alpha$ or $\beta$. Hence, the interaction with the nucleus induces on the electron frequency a splitting into two lines of similar intensity. The frequency difference between the two lines (given in units of Tesla or Hertz) is the HFC. The term "constant" is used although it is a factor. Figure 11 shows a special situation in which the three-spin system R$_1$H + R$_2$ might operate: The half of $\Delta g$ is close to the value of the HFC. In this case, half of the population of R$_1$ has a Larmor frequency close to that of R$_2$, while the other half is far off that matching frequency. Hence, in one half of the radical pair, the total spin-state remains, while it is oscillating in the other half. If a radical pair is born in the triplet $T_0$ state, half of the population will remain in the triplet state, while the other half soon will undergo interconversion to the singlet state and react. The control over the spin-dynamics, therefore, is given by the spin-state of the nucleus. This control is also called „spin-sorting" and considered to be the first step, the initial spin-physical step of the RPM.

Radical pairs in their singlet state are allowed to recombine to the recombination products (Figure 7). That reaction is spin-forbidden for radical pairs in their triplet state. In solution state, the triplet radical pair will diffuse apart and form two independent radicals surrounded by their own solvation shell each. These radicals of the so-called escape reaction might undergo subsequent chemical reactions for example with solvent molecules. If the products of the both reaction pathways are chemically distinguished, *i.e.*, have different chemical shifts, they will appear in an NMR spectrum with opposite sign as either emissive (negative) or enhanced absorptive (positive) signals. The nuclear polarization pattern appearing in NMR shows positive ("enhanced absorptive") and negative ("emissive") signals. Hence, spin-sorting can be observed by NMR if the products are different chemical species allowing for the second, the spin-chemical step of the RPM. Therefore, photo-CIDNP NMR spectra transiently show intensity patterns having the same area of positive and negative signal intensities. Photo-CIDNP NMR provides indeed an attractive hyperpolarization technique since it relies simply on the irradiation of the sample by visible light.

The efficiency of photo-CIDNP mechanisms is highly dependent [126-128] on the strength of the magnetic fields. Photo-CIDNP can also appear [126, 128-129] at low fields comparable to HFCs, at the earth field, as well as under solid-state conditions [130-135]. In confined systems, such as SCRP is micelles or biradicals, CIDNP can also exhibit features [136-138] caused by the electronic exchange coupling found at $B_0 = 2\langle J_{ex}\rangle$. In solids, CIDNP is strongly affected [133, 139-140] by non-averaged spin interactions, such as anisotropic HFC and electron-electron dipolar coupling. Generally, the spin dynamics underlying CIDNP in liquids and in solids is considerably different. For these cases, we refer to the literature [123-124, 132, 141].

## 7. Optical nuclear polarization (ONP)
For Spin-Chemistry, transient magnetic species are required. In most cases, SCRPs are discussed. A second species of relevance are molecular triplet states the substates of which are formed and decay with different kinetics. In 1967, Maier et al. observed [142] nuclear hyperpolarization in pure anthracene single crystals upon illumination with unpolarized white light. Later the effect was also found in doped crystals of other condensed aromatic compounds. The effect has a maximum at rather low fields (around 0.01 T), although it often remains observable at higher fields. Mostly samples are polarized outside the magnet and, due to long $T_1$, transferred conveniently into the magnet for the NMR measurement.

It has been shown that the origin of the optical nuclear polarization (ONP) [143-144] is electron hyperpolarization, called optical electron polarization (OEP), occurring in the triplet state. Upon light excitation, excited electronic singlet states are populated (Figure 12). ISC allows for population of electronic triplet states. Owing to zero-field splitting (ZFS), in extended molecules the three triplet electron states are not degenerate. As demonstrated by Wolf et al. [142] and Schmidt and van der Waals [145-146] in the late 1960s, the spin-orbit coupling pathways allowing for ISC are highly spin-selective. Therefore, often only one of these three electronic triplet states is populated – the one that has the symmetry matching the preceding excited singlet state. Similarly, the decay to the ground state $S_0$ is controlled by spin-selective ISC. Hence, non-Boltzmann electron population and, therefore, electron hyperpolarization is created. This electron hyperpolarization can be transferred to nuclear hyperpolarization by static hyperfine coupling during the lifetime of the triplet state. As shown by both Veeman et al. [147] and Stehlik et al. [148-150], the optimum polarization transfer to nuclei can be reached if a selective mixing of the electronic triplet states by a hyperfine interaction occurs, which leads to level anti-crossing by conserving the total magnetic quantum number. In molecular crystals, migration of triplet excitations and their trapping may control the kinetics and site of the occurrence of ONP. Triplet-state OEP and ONP, despite using a different photo-cycle, can also be generated in negatively charged nitrogen vacancy (NV⁻) centers in diamond crystals which are presently a hot field of research [151-154].

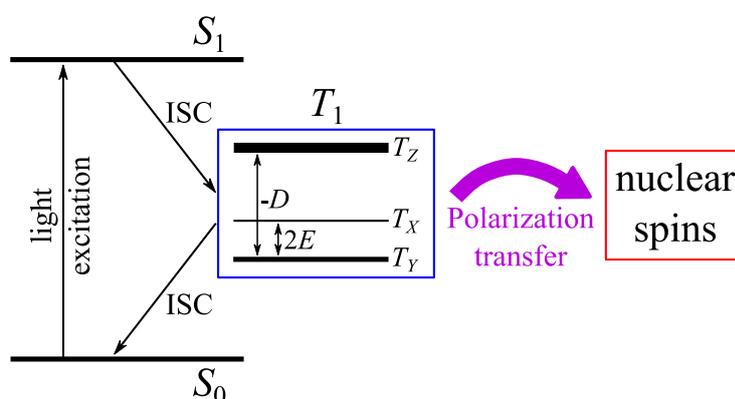

**Figure 12**. Scheme of triplet-state OEP and ONP formation. In anisotropic molecules the ISC process $S_1 \rightarrow T_1$ has a different rate for the three triplet states resulting in OEP of the $T_1$-state; the triplet states, $T_X, T_Y, T_Z$, are indicated as well as the ZFS parameters, $D$ and $E$. Alternatively, OEP can be formed due to the different $T_1$ depopulation rates in the ISC process $T_1 \rightarrow S_0$. ONP is formed by spin polarization transfer from the polarized $T_1$ state to nuclear spins of the matrix.

Hence, Spin-Chemistry deals with transient magnetic species as SCRPs and molecular triplet states interacting with the external magnetic field. Their production in chemical reactions often leads to spin-hyperpolarization which can be observed with NMR and EPR spectroscopy.

## 8. Conclusions for Spintronics

Magnetic field sensitive techniques are a well-established field for spin dependent reactions in chemistry, mainly focusing on molecules in solution phase. More recently, spin sensitive techniques have also been proposed for investigation of spin dynamics in organic photovoltaics and light emitting diodes. Based on solid films of molecules, a complete independent nomenclature has been developed. This begins with the terminology of spins, "$\alpha$" and "$\beta$" versus "up" and "down" and reaches through the entire field. An adequate exchange of knowledge between both fields is cumbersome and ends far too often at language barriers.

If one is familiar with both fields, it becomes evident that certain spin sensitive techniques have been re-invented during the last decades. This obviously can be seen as an unpleasant development: Time and money is lost to recreate already available knowledge, existing measurement setups are unused and early stage scientists do not receive their deserved recognition. Therefore, it is wise to be aware of the two fields, Spin-Chemistry and Spintronics. Examples of independently developed methods are, e.g., RYDMR for Spin-Chemisty and EDMR for Spintronics. Another example is given by CIDEP and certain EPR measurements. However, there are still techniques like CIDNP available which are not yet used in device physics.

Organic spintronics is a wide field, covering devices with spin polarized injection or light emitting diodes, which also show a magnetic field effect. We consider the relation to Spin Chemistry from the perspective of organic photovoltaics, as an example with close similarity to Spin Chemistry, in which the influence of spin-dynamics for the working solar cell is still only superficially investigated. Spin-Chemistry makes clear that singlet-triplet interaction will have an effect. Recombination of anions and cations, i.e., polarons, is a desired effect in light emitting diodes and an undesired one for solar cells. If recombination could be engineered by spin-statistics, an increase of the quantum and power conversion efficiency should be the consequence.

Concerning Tables 1 to 3 it becomes evident, that scientists from Spin-Chemistry will have a problem reading literature from Spintronics and *vice versa*. We therefore hope to give an impetus to both communities to exchange their knowledge. If successful, a foreseeable time and money consuming process will be abridged, introducing hitherto unknown techniques as CIDNP to Spintronics and the solid state systems to Spin-Chemistry. Considering the current progresses in both fields, this might be the right time to merge them together. Exchange of ideas between the two fields would strongly enhance progress in both research directions.

**List of acronyms**
CIDEP = Chemically Induced Dynamic Nuclear Polarization, CIDNP = Chemically Induced Dynamic Electron Polarization, CSS = Charge Separated State, CTS = Charge Transfer State, EDMR = Electrically Detected Magnetic Resonance, EPR/ESR = Electron Paramagnetic Resonance/Electron Spin Resonance, et and ebt = electron transfer and electron back transfer, F-pair and G-pair = free pair and geminate pair, ISC = Inter System Crossing, LFE = Low-Field Effect, MARY = Magnetically Affected Reaction Yield, MFE = Magnetic field Effect, MIE = Magnetic Isotope Effect, NMR = Nuclear Magnetic Resonance, NV = nitrogen vacancy, OMAR = Organic Magneto-Resistance, ODMR = Optically Detected Magnetic Resonance, OEP = Optical electron Polarization, ONP = Optical Nuclear Polarization, OSC = Organic Solar Cells, RPM = Radical Pair Mechanism, RYDMR = Reaction Yield Detected Magnetic Resonance, SCRP = Spin-Correlated Radical Pair, TTA = Triplet-Triplet Annihilation, ZFS = Zero Field Splitting

**Acknowledgements**
This article is dedicated to Herrn Prof. Dr. Dr. h.c. Lothar Beyer (Universität Leipzig) at the occasion of his 80$^{th}$ birthday. J.M. acknowledges the Deutsche Forschungsgemeinschaft DFG for kind support (MA 4972/2-1, MA 4972/5-1). K.L.I. acknowledges the Russian Science Foundation (grant No. 15-13-20035) and FASO of RF (project No. 0333-2016-0001). C.D. acknowledges support through the European Union's Horizon 2020 research and innovation programme under the Marie Sklodowska Curie grant agreement No. 722651, and thanks the SEPOMO project partners for interesting discussions.